\documentclass[preprint,final]{elsarticle}
\usepackage{amsmath}
\usepackage{amsfonts}
\usepackage{gensymb}
\usepackage{lineno,hyperref}
\usepackage[dvipsnames]{xcolor}
\usepackage{subcaption}
\usepackage{xcolor}
\usepackage[utf8x]{inputenc}
\usepackage{chngcntr}
\counterwithout{figure}{subsection}

\usepackage[final]{changes}
\makeatletter
\@namedef{Changes@AuthorColor}{green}
\colorlet{Changes@Color}{green}
\makeatother
\usepackage{float}
\usepackage{mathtools}
\usepackage[linesnumbered, ruled, vlined]{algorithm2e}

\journal{Nuclear Inst. and Methods in Physics Research, A}

\newcommand{\bfx}{\textbf{x}}

\newcommand{\bfw}{\textbf{w}}
\newcommand{\bfy}{\textbf{y}}

\newcommand{\bfA}{\textbf{A}}

\newcommand{\transp}{^T}

\makeatletter
\def\ps@pprintTitle{
 \def\@oddfoot{}%
 \let\@evenfoot\@oddfoot}
\makeatother

\begin{document}


\begin{frontmatter}

\title{Effect of \replaced{natural gamma background radiation}{environmental background radiation} on portal monitor radioisotope unmixing}


\author[mymainaddress]{M. Weiss}


\author[mymainaddress]{M. Fang}
\author[mymainaddress1]{Y. Altmann}
\author[mymainaddress2]{M. G. Paff}
\author[mymainaddress]{A. Di Fulvio\corref{mycorrespondingauthor}}
\cortext[mycorrespondingauthor]{Corresponding author. Tel.: +1 (217) 300-3739 }
\ead{difulvio@illinois.edu}
\address[mymainaddress]{Department of Nuclear, Plasma, and Radiological
                        Engineering, University of Illinois, Urbana-Champaign,
                        104 South Wright Street, Urbana, IL 61801, United
                        States}
\address[mymainaddress1]{School of Engineering and Physical Sciences, Heriot-Watt University, Riccarton, Edinburgh, EH14 4AS, United Kingdom}
\address[mymainaddress2]{Los Alamos National Laboratory P.O. Box 1663 Los Alamos, NM 87545}

\begin{abstract}
    National security relies on several layers of protection. One of the most important is the traffic control at borders and ports that exploits Radiation Portal Monitors (RPMs) to detect and deter potential smuggling attempts. Most portal monitors rely on plastic scintillators to detect gamma rays. Despite their poor energy resolution, their cost effectiveness and the possibility of growing them in large sizes makes them the gamma-ray detector of choice in RPMs. Unmixing algorithms applied to organic scintillator spectra can be used to reliably identify the \added{bare and unshielded }radionuclides that triggered an alarm, even with fewer than 1,000 detected counts and in the presence of two or three nuclides at the same time. In this work, we experimentally studied the robustness of a state-of-the-art unmixing algorithm to different radiation background spectra, due to varying atmospheric conditions, in the 16~\degree C to 28~\degree C temperature range. In the presence of background, the algorithm is able to identify the nuclides present in unknown radionuclide mixtures of three nuclides, when at least 1,000 counts from the sources are detected. With fewer counts available, we found larger differences of approximately 35.9$\%$ between estimated nuclide fractions and actual ones. In these low count rate regimes, the uncertainty associated by our algorithm with the identified fractions could be an additional valuable tool to determine whether the identification is reliable or a longer measurement to increase the signal-to-noise ratio is needed. Moreover, the algorithm identification performances are consistent throughout different data sets, with negligible differences in the presence of background types of different intensity and spectral shape. 
\end{abstract}


\end{frontmatter}


\section{Background and Motivation}
One of the greatest lines of defense to deter, detect, and interdict the illicit movement of special nuclear material is the application of radiation portal monitors (RPMs) at the country's many ports of entry~\cite{Kouzes2003}.\replaced{ RPMs are capable of detecting statistically significant increases in radiation above natural background. Anything at activities close to or below natural background has zero chance of being picked up by RPMs}{ The RPMs use both gamma~ray and neutron detection to sense the presence of radioactive materials.}. RPMs typically encompass \replaced{plastic}{organic} scintillators, e.g., polyvinyl-toluene (PVT), sensitive to gamma rays, and He-3 proportional counters\added{ or $^6$LiZnS based scintillators}, sensitive to neutrons~\cite{RPM}.\added{ For most vehicle and rail RPMs, the use of plastic scintillator allows to cover very large solid angle at reasonable costs, thus achieving high total efficiency; for pedestrian RPMs (much smaller pillar spacing), NaI(Tl) or other spectroscopy-capable detector materials are typically used in newer RPM models.} The \replaced{prompt detection}{accurate identification} of radioactive sources must be performed in a short time window of a few seconds to keep traffic flowing properly.
The performance of a portal monitor in terms of sensitivity, i.e., maximization of the positive detection rate, depends on the detection efficiency of the system and its form factor, which should be optimized for a specific application~\cite{Paff2016}. 
In previous studies~\cite{Altmann2020}, we experimentally demonstrated the use of a sparsity-promoting Bayesian algorithm capable of unmixing the signatures from weak gamma-ray sources, detected by organic scintillators. Our algorithm, hereafter referred to as the unmixing algorithm, allowed to identify radioactive sources based on measured spectra consisting of less than 500 counts despite the relatively low energy resolution featured by organic scintillators. In an unknown spectrum with approximately 1000 counts, the algorithm is able to identify up to three gamma~emitting radionuclides, and a few hundred of counts from weapons grade plutonium results in an alarm rate of 80\%~\cite{Altmann2020}.
The algorithm relies on a pre-compiled library of radionuclides to correctly identify the mixture components. The library encompassed the most common gamma-ray emitting nuclides among naturally-occurring radioactive material (NORM), laboratory gamma-ray sources, radio-isotopes for medical applications, and special nuclear materials.
However, in our previous analysis, the library lacked consideration for the presence of a time-dependent radiation background, as well as the likely case where the radioactive material is being shielded by a variety of different materials.
The presence of background radiation can have a large impact on the detection sites, with cosmic and terrestrial radiation being seen on the detected spectra.
In the inherently low signal-to-background measurement conditions at RPMs, the intrinsic observation noise should not be neglected and can be modeled as Poisson noise, i.e., shot noise. 
In this study, we focus on extending the use of the sparsity-promoting Bayesian algorithm to identify multiple radionuclide sources in an unknown mixture in the presence of different background conditions, while shielding scenarios are left for future work. 
\replaced{Changes}{Increase} in the background count rate could be the effect of three different causes: (1) the actual increase of the background radiation, (2) a detector-dependent readout increment\added{, or (3) background shielding by large vehicles close to the RPM (also known as \textit{ship-effect}~\cite{KOUZES200889})}.
The \replaced{first cause}{former} can be due to rainfall, which scavenges the radon progeny within clouds and causes its deposition on the ground~\cite{Livesay2014}.
The \replaced{second cause}{latter} can be determined by several factors, such as a decreased gain of the detector photomultiplier tube (PMT) due to an increase of the external temperature~\cite{Wang2014,Gehman2007,Liebson1950}. The gain decrease may result in an increase of the energy range corresponding to the selected detection window, and therefore in an increase of the overall count rate. In general, the temperature has multiple effects on a PMT operation, including dynode gain change and cathode sensitivity~\cite{HamamatsuPhotonics2007}. 
We recorded the detected background radiation and the environmental temperature to understand their correlation, and compensated for detector dependent effects, if necessary.

\section{Methods}
This section describes the experimental methods to measure temperature, humidity, and background radiation, build a reference radionuclide library, and test the unmixing algorithm under different background conditions.

\subsection{Experimental Methods}
\begin{figure}[H]
\captionsetup{font=footnotesize}
\centering
\begin{subfigure}[H]{0.49\textwidth}
  \includegraphics[width=\textwidth]{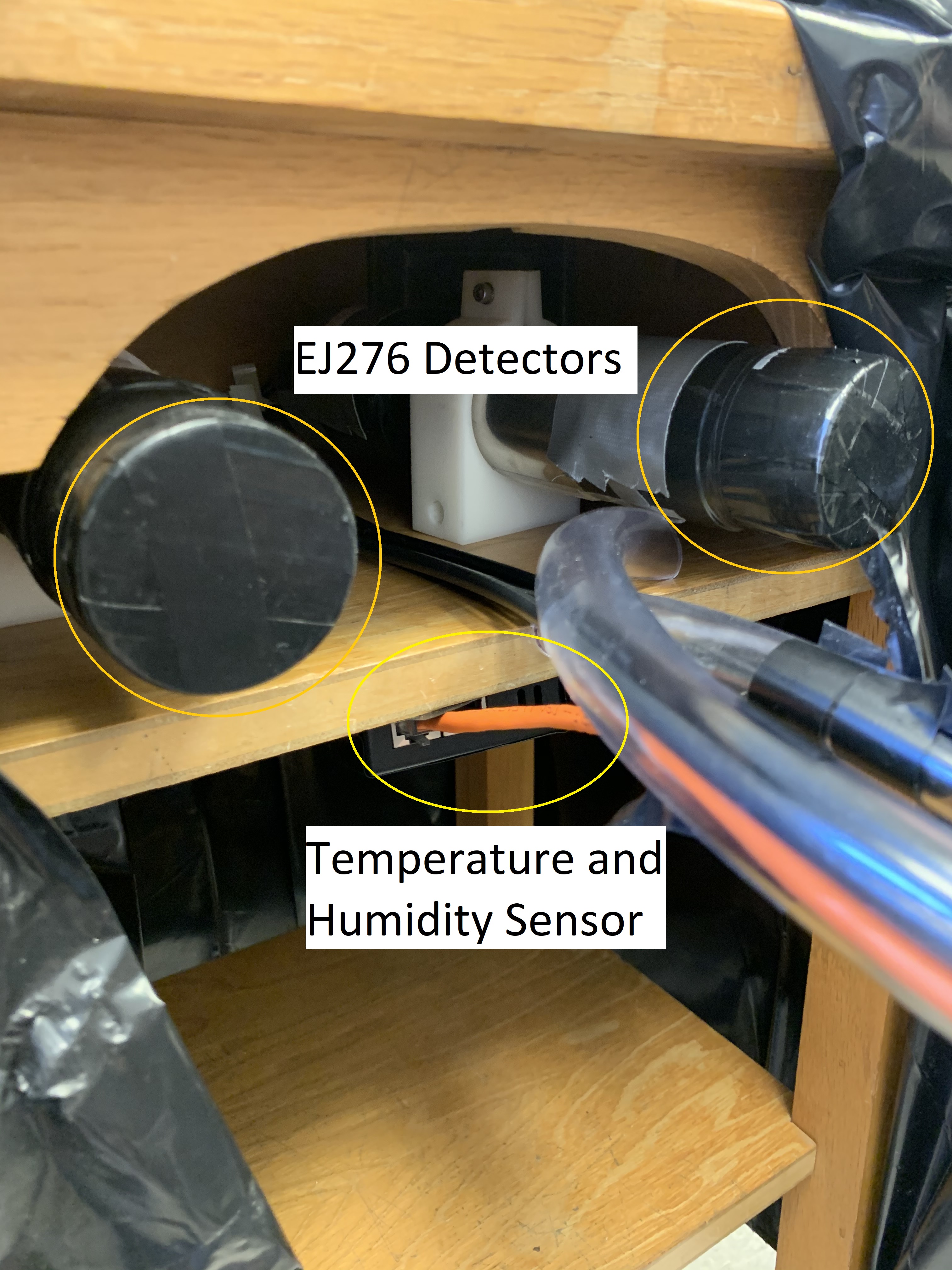}
  \caption{Cabinet detector setup: two EJ-276 detectors with PMTs, temperature and humidity sensor attached to underside of shelf}
  \label{fig:inside}
\end{subfigure}
\begin{subfigure}[H]{0.49\textwidth}
  \includegraphics[width=\textwidth]{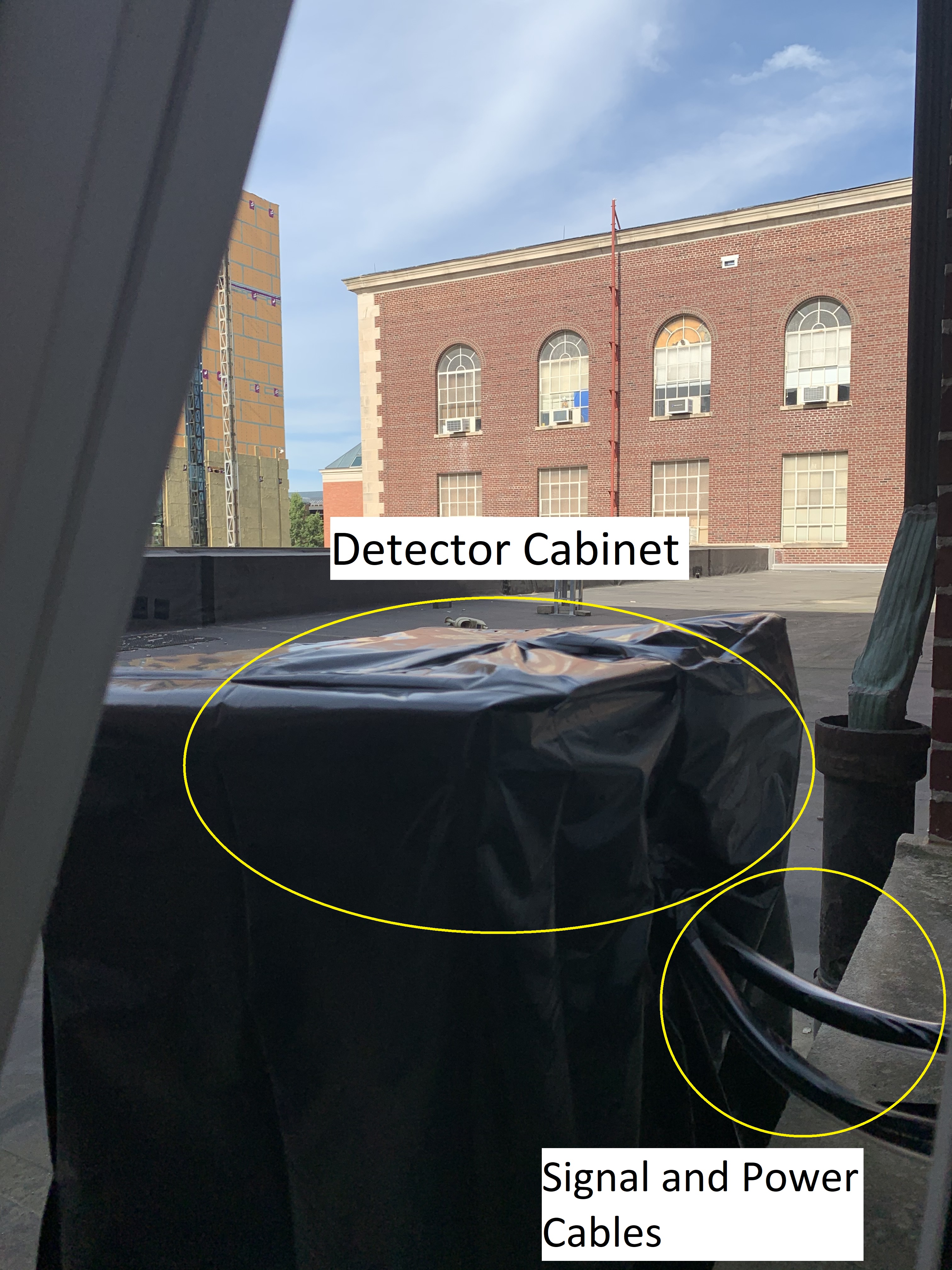}
  \caption{Outdoor cabinet setup, tubes containing connecting wires to the digitizer and data logger indoors}
  \label{fig:outside}
\end{subfigure}
  \caption{Experimental setup}
  \label{fig:expsetup}
\end{figure}

 Background radiation, humidity, and temperature measurements were taken from an outdoor environment in Urbana, Illinois, US ($40.1106^{\circ}$~N, $88.2073^{\circ}$~W). The detectors for the long-term radiation, temperature and humidity data gathering were housed in an outdoor shelving unit with the sides covered by a thin sheet of plastic to keep out rainfall, as shown in Fig.~\ref{fig:expsetup}. This approach prevented the inside of the shelving unit from becoming a micro-environment with a possibly different temperature and humidity than the outside. The detector and PMT were in thermal equilibrium with the environment, therefore, it was possible to study the correlation between the measured temperature and the background count rate, directly. We did not include a temperature stabilization system, as done in some RPMs deployed in the field, because our objective was to investigate the effect of the temperature fluctuations in a relatively small range around room temperature on the readout. 
 

 We measured the background counts using a pair of \added{5.08~cm~$\times$~5.08~cm }{EJ-276} organic scintillators by Eljen Technologies. EJ-276 (4.546$\times$10$^{22}$ and 4.906$\times$10$^{22}$, hydrogen and carbon atoms per cm$^3$, respectively) is a relatively new type of plastic scintillator, with pulse shape discrimination capability~\cite{EljenTechnology}. A 9214B PMT by Electron tubes converted the radiation-induced light pulses into current pulses. According to the manufacturer, the temperature coefficient of the PMT due to change in photocathode sensitivity and electron multiplier gain is -0.3$\%$/\degree C~\cite{Hamamatsu2}. The PMTs were connected to a desktop high-voltage power supply (CAEN DT5533EN) that supplied -1950 V to the first detector and -~1838~V to the second one to gain match both the detectors at a pulse integral of 5.15~V~ns/dt, corresponding to the Compton edge of a $^{137}$Cs source, selected as the  as the pulse integral corresponding to the 80\% of the Compton edge with respect to its maximum. The lower detection threshold was 15 keVee. 
 The detected pulses were digitized by a 14-bit 500-MSps CAEN DT5730 digitizer and acquired by the CAEN CoMPASS software~\cite{caencompass}. Raw data were then processed using a custom software written in Matlab 2017, The MatWorks, Inc. Temperature and humidity measurements were taken outdoor using a Teracom humidity and temperature sensor (TSH300v2) connected to a Teracom Ethernet data logger (TCW210-TH), which was then connected via Ethernet to a lab computer for data storage. The data logger stored temperature and humidity measurements in 5 minute intervals. 
 
 
 
\subsection{Radioisotope Mixtures and Library}
\begin{figure}[H]
\captionsetup{font=footnotesize}
\centering
\begin{subfigure}[H]{0.7\textwidth}

  \includegraphics[width=\textwidth]{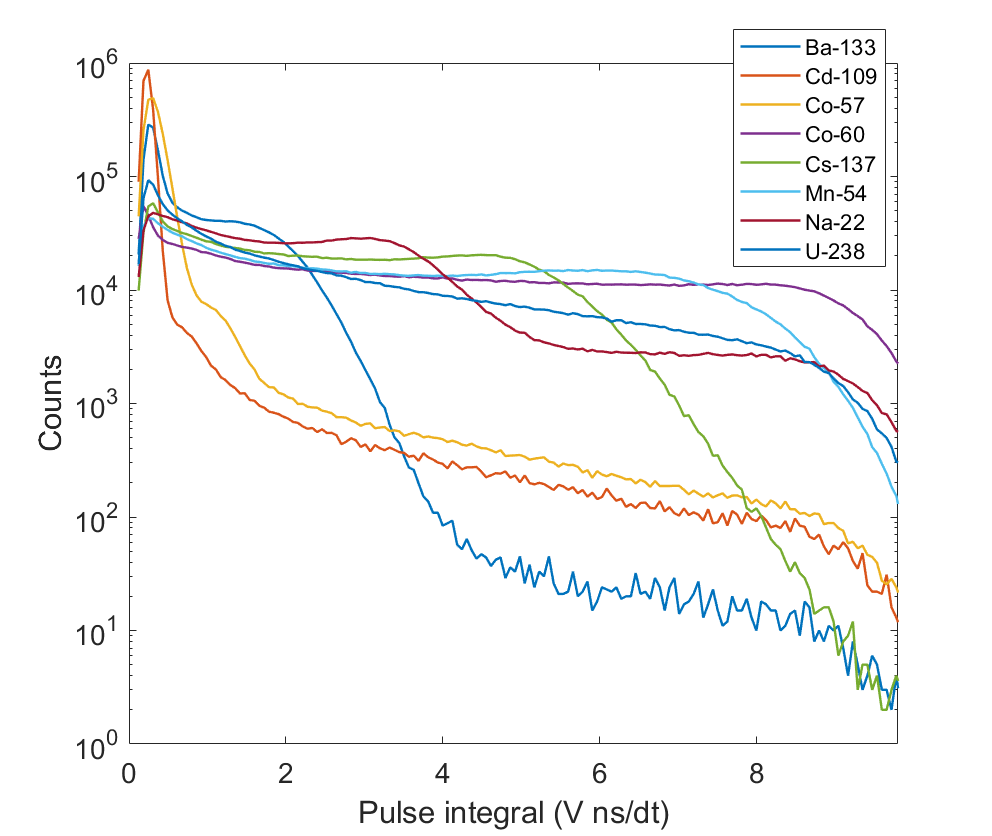}
  \caption{Pulse integral distribution of the measured counts}
  \label{fig:pulseI}
\end{subfigure}
\begin{subfigure}[H]{0.7\textwidth}
  \includegraphics[width=\textwidth]{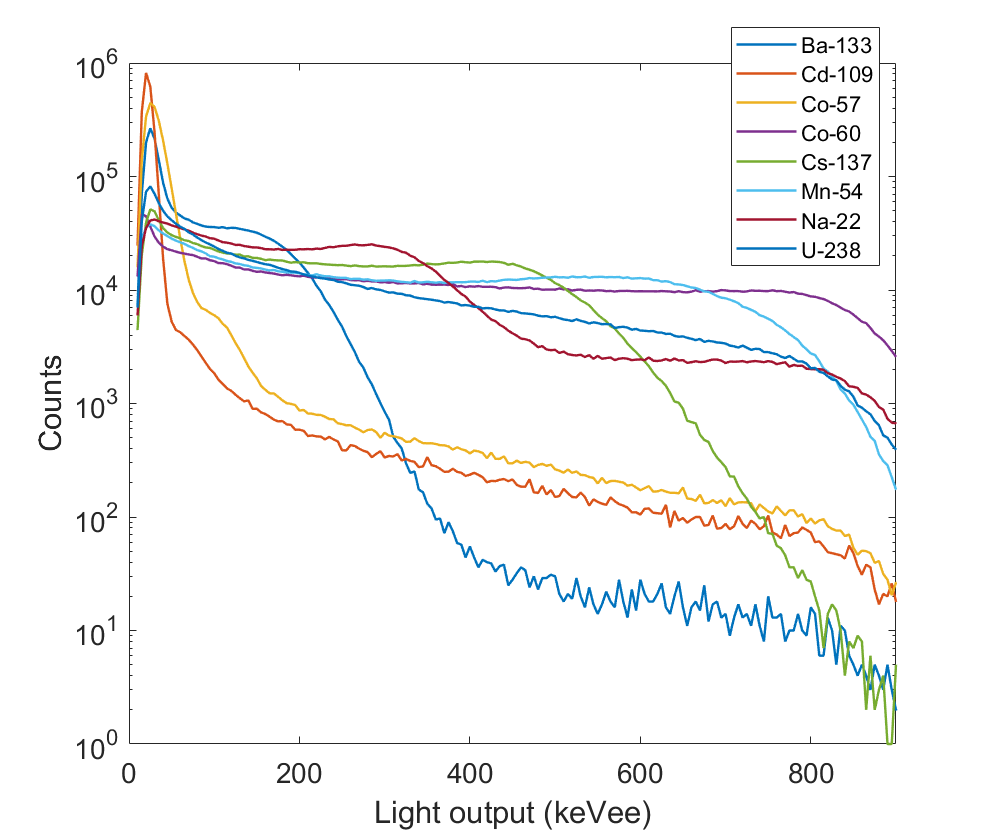}
  \caption{Light output distribution}
  \label{fig:pulseLO}
\end{subfigure}
  \caption{Library of radionuclides, encompassing eight gamma-ray emitting isotopes.}
  \label{fig:Library}
\end{figure}

We generated a variety of radioisotope mixtures to test the effectiveness of the unmixing algorithm. The algorithm performance in identifying the present radionuclides and correctly estimating their fractions were calculated, as a function of the number of detected counts. 
The library spectra were measured using the same experimental setup used for the background measurement. In this case, the acquisition was performed indoors. The isotope library included: $^{133}$Ba, $^{109}$Cd, $^{57}$Co, $^{60}$Co, $^{137}$Cs, $^{54}$Mn, $^{22}$Na, and $^{238}$U. Each source, except $^{238}$U, had an activity of 1~$\mu$Ci\added{ (3.7$\times$10$^{4}$~Bq)} as of October 19, 2017. The $^{238}$U source was in the shape of a hollow rod coated by a 1.07-mm thick aluminum liner (2.54~cm outer diameter, 1 cm inner diameter and 25 cm length) of 1.8~Ci\added{ (6.66$\times$10$^{10}$~Bq)} activity, measured on October 16, 2013. The measured spectra are displayed in Fig.~\ref{fig:Library}. Each isotope was measured over a different time period, enough to collect approximately 2,000,000 counts, in total. The light output was calibrated in electron-equivalent units~\cite{Akimov2002}, using the Cs-137 Compton edge pulse integral value.\added{ The background of indoor measurements was not removed due its negligible scale compared to the radioisotope spectra counts.}
With these libraries created, the mixtures were then generated and used as input of the unmixing algorithm. 
 
We generated six different mixtures using the pulse integral spectra information from the library. The mixture isotope amounts are listed in Table~\ref{table:Mixtures}. These mixtures were chosen based on naturally occurring isotope combinations. Spectra with similar shapes but including different nuclides were also chosen to test the algorithm spectral sensitivity.
Each mixture generation began by linearly combining the spectra corresponding to the selected radionuclides in the proportions shown in Table \ref{table:Mixtures}.

\begin{table}[H]
\centering
\caption{True isotope proportions of the synthesized mixtures}
\resizebox{12cm}{!}{
\begin{tabular}{|l|r|r|r|r|r|r|} 
\hline
       & \multicolumn{1}{l|}{Mixture \#1} & \multicolumn{1}{l|}{Mixture \#2} & \multicolumn{1}{l|}{Mixture \#3} & \multicolumn{1}{l|}{Mixture \#4} & \multicolumn{1}{l|}{Mixture \#5} & \multicolumn{1}{l|}{Mixture \#6}  \\ 
\hline
Ba-133 & 0                                & 0                                & 0                                & 0                                & 0.33                             & 0                                 \\ 
\hline
Cd-109 & 0.34                             & 0.33                             & 0                                & 0                                & 0.33                             & 0                                 \\ 
\hline
Co-57  & 0.33                             & 0                                & 0.33                             & 0                                & 0                                & 0                                 \\ 
\hline
Co-60  & 0.33                             & 0.33                             & 0                                & 0                                & 0                                & 0                                 \\ 
\hline
Cs-137 & 0                                & 0                                & 0.33                             & 0.5                              & 0.34                             & 0.25                              \\ 
\hline
Mn-54  & 0                                & 0                                & 0.34                             & 0                                & 0                                & 0.25                              \\ 
\hline
Na-22  & 0                                & 0.34                             & 0                                & 0                                & 0                                & 0                                 \\ 
\hline
U-238  & 0                                & 0                                & 0                                & 0.5                              & 0                                & 0.5                               \\
\hline
\end{tabular}
}
\label{table:Mixtures}
\end{table}
 
 Once each mixture was generated, a set amount of stable reference background measurement, conservatively equivalent to one minute of detection time, was added to each mixture. Lastly, Poisson noise was added to each mixture to simulate Poisson noise in both the isotopes and background radiation.
 The response from the two detectors was comparable, therefore, results are only shown from one detector. 
 Each mixture was generated 100 times, with background and Poisson noise applied each time, and the unmixing algorithm was run over each iteration of the mixtures. The average value for predicted radionuclide relative amounts and probability of appearance was calculated from the unmixing algorithm iterations and reported in the heat maps in Fig. \ref{fig:heatmaps} and the RMSE graphs in Fig. \ref{fig:rmse}. This approach was then repeated for each of the varied amount of mixture counts (\added{500, }1000, 5000, 10000, 25000, 50000, and 100000 not including background or Poisson noise). The scale of typical vehicle RPM with four plastic scintillators may have a total background count rate of about 1000 gamma counts per second, with many alarms just being hundreds to thousands of gamma counts per second above background. These situations account for many of our lower count levels in the range, such as 500, 1000, and 5000 counts. The upper end of the count level range is included for better visualization, as well as including situations where higher count alarms do occur.

\subsection{Computational Methods}
As mentioned above, we used a Bayesian algorithm to find the mixing coefficients associated with each library nuclide to assess their contribution to the 
measured spectrum.
In previous studies, we analyzed several Bayesian approaches to solve this problem~\cite{Paff2018,Altmann2020}. These methods exploit the posterior distribution of the mixing coefficients, by combining the observed data with available prior information. We have observed that a sparsity-promoting approach using a Bernoulli-truncated Gaussian (BTG) prior model yielded state-of-the-art 
estimates for the nuclide fractions in the unknown mixtures~\cite{Altmann2020} and, therefore, we only consider this method in this work. \\
Given an observed light output spectrum $\bfy=[y_1,\ldots,y_M]\transp$, observed in $M$ energy bins, which is associated with a mixture of up to $N$ known sources, whose individual spectral responses are denoted by $\left\lbrace \bfA_{:,n}\right\rbrace_{n=1,\ldots,N}$ and gathered in the $M \times N$ library matrix $\bfA=[\bfA_{:,1},\ldots,\bfA_{:,N}]=[\bfA_{1,:}\transp,\ldots,\bfA_{M,:}\transp]\transp$. Each $\bfA_{m,:}$ is a row vector gathering the spectral responses of the $N$ known sources in the $m$th energy bin.
 The coefficient associated with the $n$th source, corresponding to the amount present in the mixture, is denoted by $x_n$ and the $N$ coefficients are gathered in the vector $\bfx=[x_1,\ldots,x_N]\transp$.
 In this work, $N=9$ since we included the eight library sources and a reference background spectrum, acquired over 12 hours and normalized to 1 (unit integral), like all the other library spectra.
At a first approximation, we assume a linear mixing model of the source components, which can be expressed in matrix form as $\bfy \approx \bfA \bfx$. 
$\bfA$ is known and is omitted in all the conditional distributions hereafter. As mentioned above, the observation noise is modeled by Poisson noise, leading to the following form of the likelihood
\begin{eqnarray}
\label{eq:likelihood}
f(y_m|\bfx)= \left(\bfA_{m,:}\bfx\right)^{y_m} \exp{\left[-\bfA_{m,:}\bfx\right]}/y_m!, \quad \forall m=1,\ldots, M.
\end{eqnarray}
 The entries of $\bfy$ are independently distributed, i.e., $f(\bfy|\bfx)=\prod_{m=1}^M f(y_m|\bfx)=\prod_{m=1}^M f(y_m|\bfA_{m,:} \bfx)$ and are conditioned on the value of $\bfx$. Bayesian methods rely on the knowledge of prior information available about $\bfx$, the coefficients, to enhance their recovery from the observable $\bfy$, the light output spectrum. The a-priori information is the prior distribution $f(\bfx)$ and the estimation of $\bfx$ can then be achieved using the posterior distribution $f(\bfx|\bfy)=f(\bfy|\bfx)f(\bfx)/f(\bfy)$. \\
The efficient sparsity-promoting  BTG prior model of $\bfx$ is described in Eq.  \eqref{eq:BTG_prior}
\begin{eqnarray}
\label{eq:BTG_prior}
f(x_n|w_n) = (1-w_n)\delta(x_n) + w_n\mathcal{N}_{\mathbb{R}^+}(x_n;0,\sigma_n^2), \quad \forall n=1,\ldots,N\nonumber \\
f_n(w_n=1)=\pi_n, \quad \forall n=1,\ldots,N,
\end{eqnarray}

In Eq.\eqref{eq:BTG_prior}, $\delta(\cdot)$ denotes the Dirac delta function, which is equal to 1 when $x_n=0$ and 0 elsewhere and where $\mathcal{N}_{\mathbb{R}^+}(x_n;0,\sigma^2)$ is a probability density function (p.d.f.) truncated Gaussian distribution, defined on $\mathbb{R}^+$ to enforce the non-negativity of the elements of $\textbf{x}$. The truncated Gaussian prior has hidden mean $0$ and hidden variance $\sigma^2$these are the mean and variance of the non-truncated Gaussian distribution). The presence of the $n$th source is controlled by the binary variable $w_n$, which is equal to 1 when the $n$th is present and 0 otherwise. $\pi_n$ is the prior probability of presence of the $n$th source. 

We set $\pi_n=1/N, \forall n$ as we expect a limited number of sources to be simultaneously present in the mixture, while we do not wish to promote any specific source. These parameters can however modified by the practitioners. We set the variances  $\{\sigma_n^2\}$ as in Eq. \eqref{eq:variance} for each source

\begin{eqnarray}
\label{eq:variance}
\sigma_n^2=0.1\sum_{m=1}^M y_m .
\end{eqnarray}


Instead of considering a prior model only for $\bfx$, Eq. \ref{eq:BTG_prior} defines a joint prior model for $(\bfx,\bfw)$, where $\bfw=[w_1,\ldots,w_N]\transp$, expressed as $f(\bfx,\bfw)=\prod_{n=1}^N f(x_n|w_n)f_n(w_n)$. The proposed unmixing algorithm thus aims at estimating jointly $(\bfx,\bfw)$, i.e., at performing jointly the source identification (through $\bfw$) and quantification (through $\bfx$).

Using the Bayes' rule, the joint posterior distribution of $(\bfx,\bfw)$ is given by $f(\bfx,\bfw|\bfy)=f(\bfy|\bfx)f(\bfx,\bfw)/f(\bfy)$.

The algorithm adopted in this paper and originally described in~\cite{Altmann2020} relies on approximate Bayesian estimation and builds an approximate distribution $Q(\bfx,\bfw) \approx f(\bfx,\bfw|\bfy)$ whose moments are much simpler to evaluate than those of $f(\bfx,\bfw|\bfy)$. The method belongs to the so-called class of expectation propagation (EP) methods~\cite{minka2001expectation} to provide approximate point estimates of the mean and the covariance of the posterior distribution of $\bfx$ (and $\bfw$). It offers several advantages compared to traditional approaches that exploit the posterior distribution using Hamiltonian Monte Carlo methods~\cite{Paff2018, Vehtari2014} and is also motivated by the fact that the posterior means $E_{f(\bfx,\bfw|\bfy)}[\bfx]$ and $E_{f(\bfx,\bfw|\bfy)}[\bfw]$ associated with the posterior distribution $f(\bfx,\bfw|\bfy)$ are intractable analytically. Further details on the EP algorithm and its implementation can be found in our previous work~\cite{Altmann2020} and are available online in its current version at~\cite{GitHub}.
 
\subsection{Unmixing Algorithm Classification Methodology}
Once the unmixing algorithm runs through 100 iterations of the six generated mixtures, the output consists of two sets of values: 1) Predicted amounts of each radioisotope for each of all six mixtures in all 100 cases, and 2) Predicted probability of each radioisotope's presence in the mixture for all six mixtures in all 100 cases. These sets of values allow for the creation of the heatmaps shown below in Fig. \ref{fig:heatmaps}. Once all generated mixtures at different count levels have been run through the unmixing algorithm, the average for the isotope presence probability can be calculated. The values inside the heatmaps in Fig. \ref{fig:heatmaps} represent these average probabilities of presence for each isotope at each mixture counts level. Similarly, the color bar represents this information as a color scale, with blue indicating a very low isotope presence probability and red indicating a high isotope presence probability. These probabilities allow for analysis on how the unmixing algorithm classifies each mixture as a combination of different isotopes, as well as how these probabilities change with varying counts.

\subsection{Algorithm metrics}
We compared the unmixing performance of the algorithm under various background conditions and as a function of the total number of counts using the root-mean-square error (RMSE, Eq.\ref{eq:RMSE}). 

\begin{equation}
\label{eq:RMSE}
    \mathrm{RMSE} = \sqrt{\dfrac{||\bfx-\hat{\bfx}||_2^2}{N}}
\end{equation}
between the known nuclide fractions $\bfx$ and their estimated values $\hat{\bfx}$, where $N$ is the number of nuclides in the spectral library, including one reference background spectrum.
The RMSE was calculated for each of the 100 generated mixtures, then averaged and the standard deviation was taken. The RMSE was calculated considering mixtures to which each of the four background types (see Results section) was added to the original mixture, with \emph{stable} background remaining in the library. This approach allows for evaluating how the isotope identification effectiveness changes with changing background conditions in relation to the library reference background.


\section{Results}
\subsection{Background, Temperature, and Humidity Data}
 
\begin{figure}[H]
\captionsetup{font=footnotesize}
\centering
  \includegraphics[width=1\textwidth]{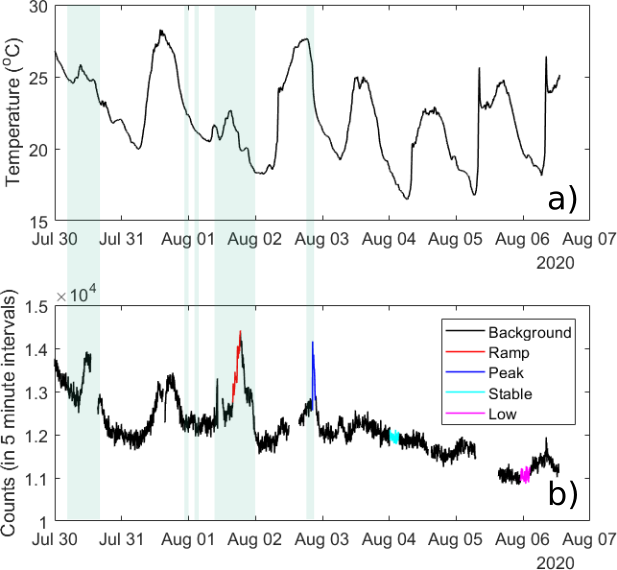}
  \caption{Outdoor sensor data over time for (a) Temperature and (b) Counts. Rainfall events occurred over the time spans represented in green.}
  \label{fig:outdoordata}
\end{figure}

\begin{figure}[H]
\captionsetup{font=footnotesize}
\centering
  \includegraphics[width=\textwidth]{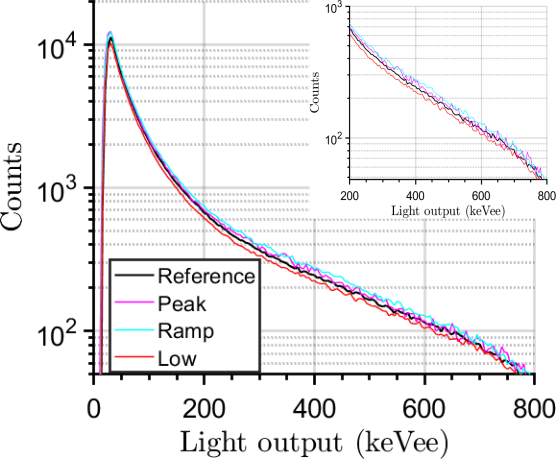}
 \caption{Spectra measured under different conditions for one hour up to 800 keVee.}
  \label{fig:bgspectra}
\end{figure}

 Fig. \ref{fig:outdoordata} displays how the counts measurements changed over one week, from July 30 to August 6, 2020. The graph shows binned counts over five-minute intervals, with a variety of trends 
 throughout the week. The large spikes in counts measurements align with rainfall events on their respective days.
 Data in selected time frames were used to create background spectra at four different background conditions, labeled in Fig. \ref{fig:outdoordata} as \emph{stable}, \emph{ramp}, \emph{peak}, and \emph{low}. The stable background was chosen at a time without any rainfall and relatively average temperature and humidity for the time of day, representing a baseline background level. Peak background corresponded with a time of heavy rainfall, creating a large spike in counts as expected. Ramp background corresponded with a time span when rain was beginning and the number of counts were rising to a peak region. Lastly, the low background region indicates a time span with the lowest number of recorded counts, also corresponding with low humidity for the time of day. These four regions are essential for analysis of how a changing background spectra and counts can interfere with the effectiveness of the unmixing algorithm.
 The main difference between the different background spectra can be noticed in Fig. \ref{fig:bgspectra}, with the \emph{low} background being characterized by an overall lower intensity throughout the whole spectrum, compared to the other background spectra. 
Extended count gaps in Fig. \ref{fig:outdoordata}(b) correspond to detector calibration periods. 
The calibration was performed by placing the Cesium-137 source in front of the detectors and checking that the shape and the Compton edge location were consistent with previous acquisitions. 

 Figures ~\ref{fig:outdoordata}(a) and (b) show how both the temperature and counts changed over the testing time period. As shown, the temperature had significant variation as a result of the varying high temperatures during the day, transitioning from night-to-day and day-to-night, and the sun's position relative to the detector. During the measurement, the temperature varied from 16.48\degree C to 28.28\degree C, with an average of 21.64\degree C. 
 

 
\begin{figure}[H]
\captionsetup{font=footnotesize}
\centering
  \includegraphics[width=\textwidth]{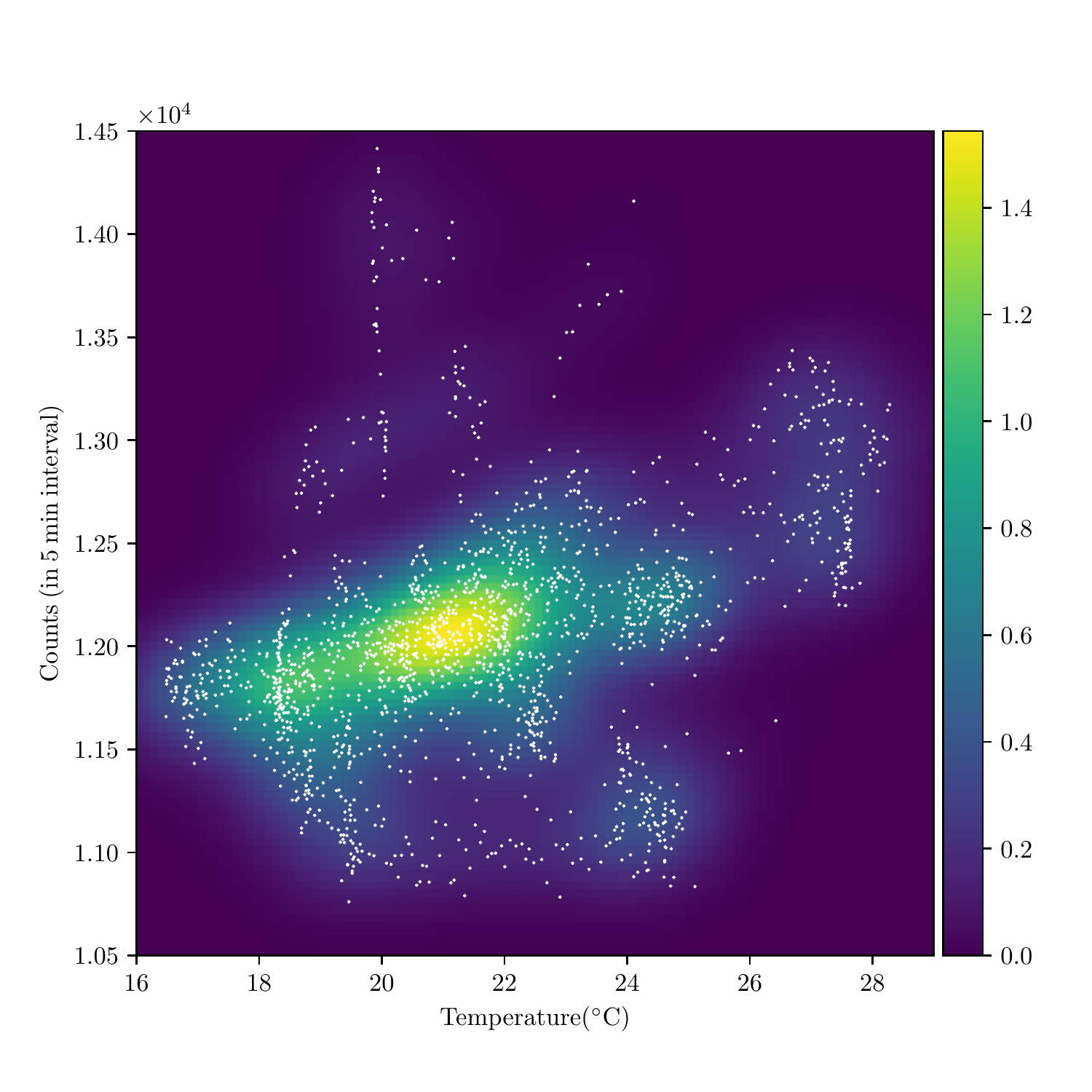}
  \caption{Counts as a function of Temperature Scatter Density Plot}
  \label{fig:CvTscatterdensity}
\end{figure}

We analyzed the dependence of the background counts, in five-minute intervals, with the temperature to identify potential gain drifts caused by temperature gradients, and determine whether further spectrum re-calibration was necessary. Fig.~\ref{fig:CvTscatterdensity} shows the scatter-density plot of the measured counts as a function of the recorded temperature. The plot intensity is the frequency of occurrence of the counts. 

The sample standard deviation for counts in five minute intervals, $\sigma$, is approximately 588.7 counts, with some anomalies in recordings created by days of heavy rainfall. These anomalous results can be analyzed on a case-by-case basis, with most being a result of short bursts of heavy rainfall. 
At a temperature slightly above 20\degree C shown in Fig.~\ref{fig:CvTscatterdensity}, a relatively high count rate was recorded due to a day of very heavy rainfall that caused an increase in radioactivity from deposition of the radon progeny on the ground. Similarly, a region of low count rates (11,000 in five minutes, on average), and temperature ranging from 19\degree C to 24\degree C, was measured when the humidity was low. 
As expected, a positive weak correlation between count rate and temperature was found.
With reference to Fig.~\ref{fig:CvTscatterdensity}, the counts increased, on average, by about 1,600 from 16\degree C to 28\degree C, i.e., the temperature range of our experiment. This increment is comparable with 3$\sigma$ of 1766.1. Therefore, we concluded that during the experiment the temperature 
negligibly affected the count rate. 
As such, all of the data collected can be used for applying to the unmixing algorithm without further spectral gain correction.


\subsection{Source Identification}

\begin{figure}[H]
\captionsetup{font=footnotesize}
\centering
\begin{subfigure}[H]{0.7\textwidth}
  \includegraphics[width=\textwidth]{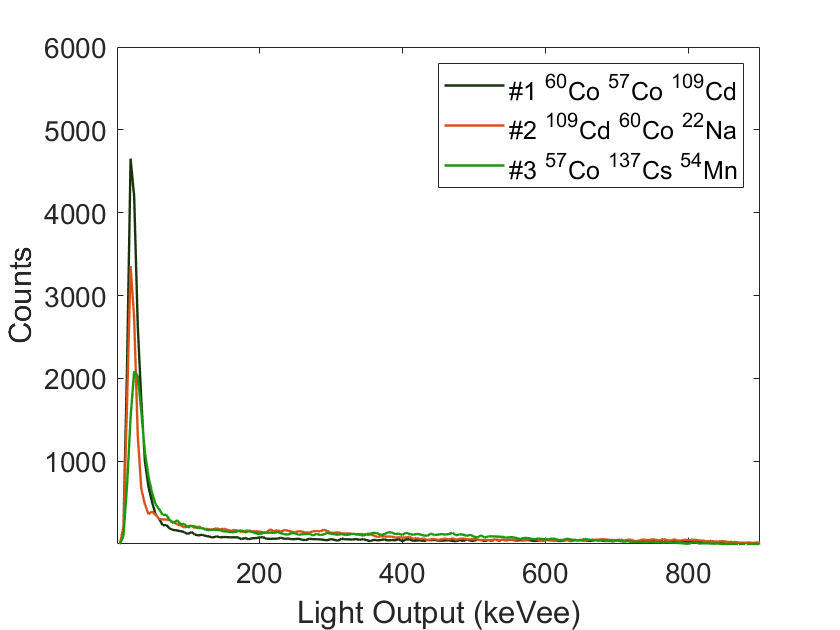}
  \caption{Isotope Mixtures 1-3}
  \label{fig:mix123}
\end{subfigure}
\begin{subfigure}[H]{0.7\textwidth}
  \includegraphics[width=\textwidth]{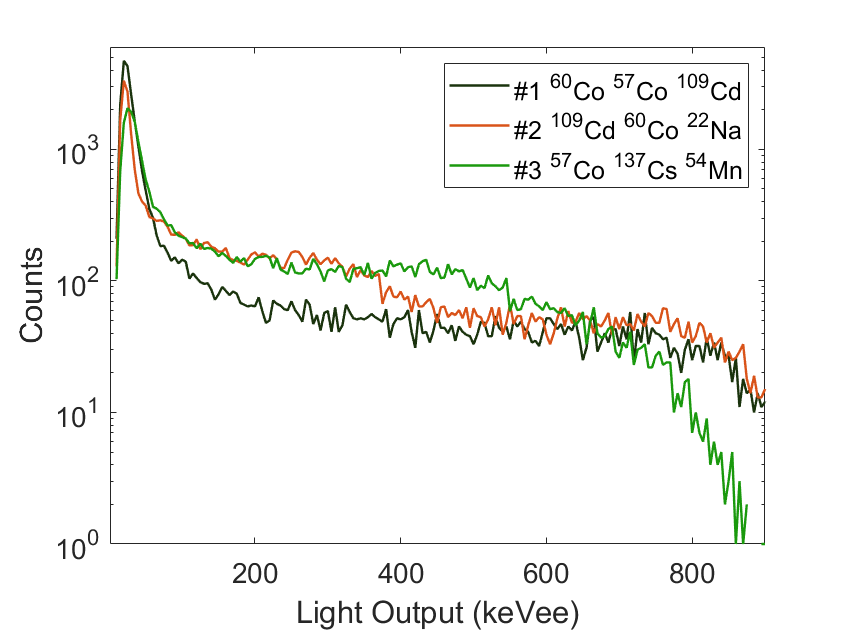}
  \caption{Isotope Mixtures 1-3 with log-scaled y-axis}
  \label{fig:mix_nolog}
\end{subfigure}
  \caption{25000 Counts Isotope Mixtures with background and Poisson noise added}
  \label{fig:mix_spectra}
\end{figure}

\begin{figure}[H]
\captionsetup{font=footnotesize}
\centering
  \includegraphics[width=1.1\textwidth]{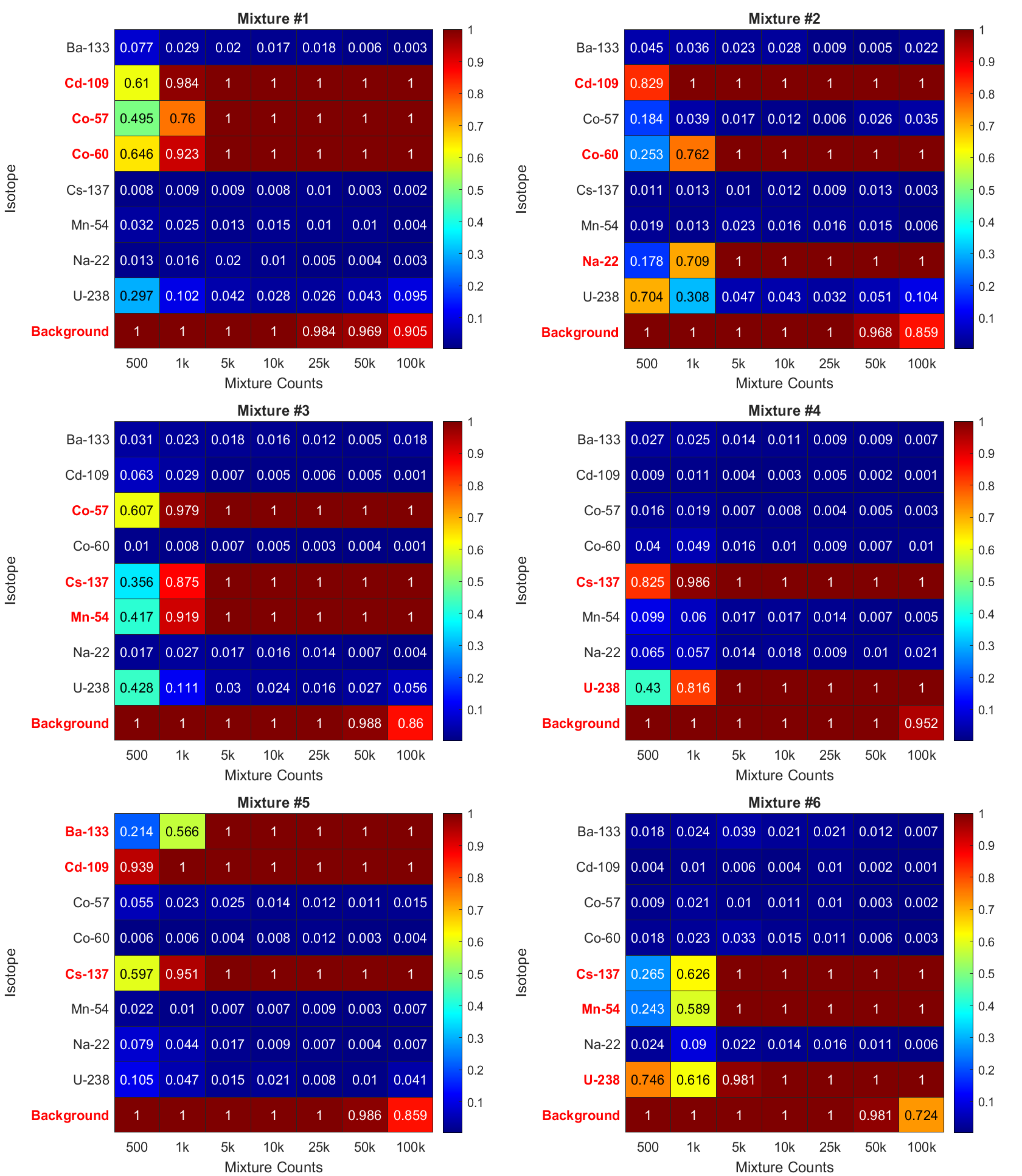}
  \caption{Heat maps of predicted probability from unmixing algorithm of isotope occurrence in six generated mixtures at varying mixture counts. The numbers on each nuclide-count combination is the predicted probability of the isotope's presence in the mixture.\added{ Isotopes in red are used to generate the mixtures.}}
  \label{fig:heatmaps}
\end{figure}

The spectra for each of the generated mixtures are shown in Fig. \ref{fig:mix_spectra}, with three spectra being shown per graph for better visualization. Each of these spectra has a total of 25,000 counts from the isotopes themselves, with additional counts being added from the reference background spectra and inserted Poisson noise.
Despite the presence of different nuclides, the spectra exhibit similar shapes, with a main peak at low light output values, and weak differences above 200~keVee. The heatmaps displayed in Fig. \ref{fig:heatmaps} show how effective the unmixing algorithm is at identifying the isotopes in each of the generated numbered mixtures. 

As seen in the six heat maps, the lowest number of counts (500) often corresponded with the predictions with the least level of certainty. 
It should be noted that the number of background counts added to the mixtures is the same, regardless of the total mixture counts. This higher relative background fraction may cause incorrect nuclide identification at low counts.
This effect can be seen in mixture \#2, which contains $^{109}$Cd, $^{60}$Co, and $^{22}$Na. At 500 counts, an incorrect 70.4\% chance for $^{238}$U to be present in the mixture is reported, with this percentage steeply dropping off as the number of counts increased.  Similarly, this is the reason why background is detected with a high probability at low number of counts, but the probability begins to decline at high count levels. 
Including the background in the potential $N$ spectral components allowed for correct identification also at low counts as shown in Fig. \ref{fig:heatmaps}. 

\begin{figure}[H]
\captionsetup{font=footnotesize}
\centering
  \includegraphics[width=0.8\textwidth]{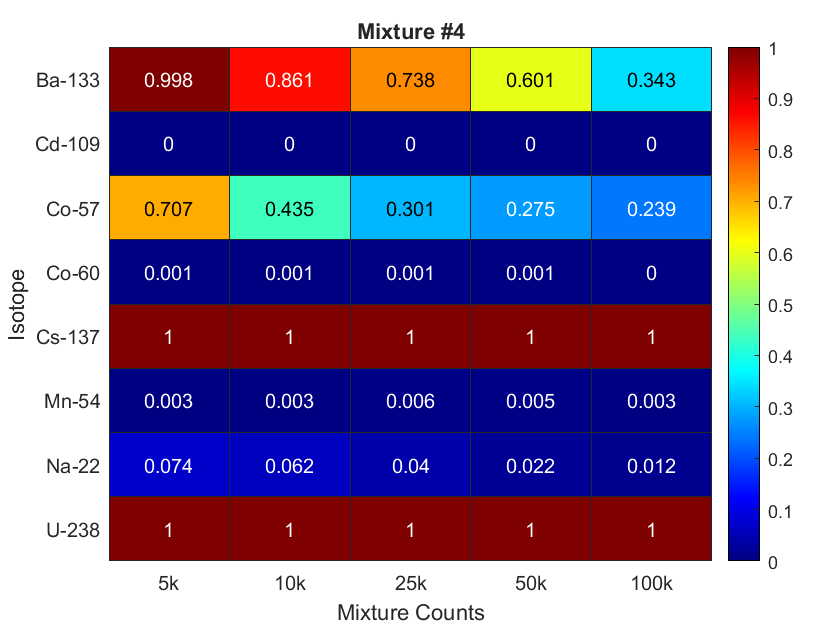}
  \caption{Heat map of the predicted probability of isotope occurrence in Mixture \#4 at varying mixture counts without background included in the library.}
  \label{fig:heatmap_bad}
\end{figure}

The effect of not including the background spectrum in the unmixing library is illustrated in Fig. \ref{fig:heatmap_bad}. In mixture \#4, encompassing only $^{137}$Cs and $^{238}$U, $^{57}$Co and $^{133}$Ba are always incorrectly included in the mixture, even when a high number of counts is available, corresponding to a low observation noise. 

\begin{figure}[H]
\captionsetup{font=footnotesize}
\centering
\begin{subfigure}[H]{0.49\textwidth}
  \includegraphics[width=\textwidth]{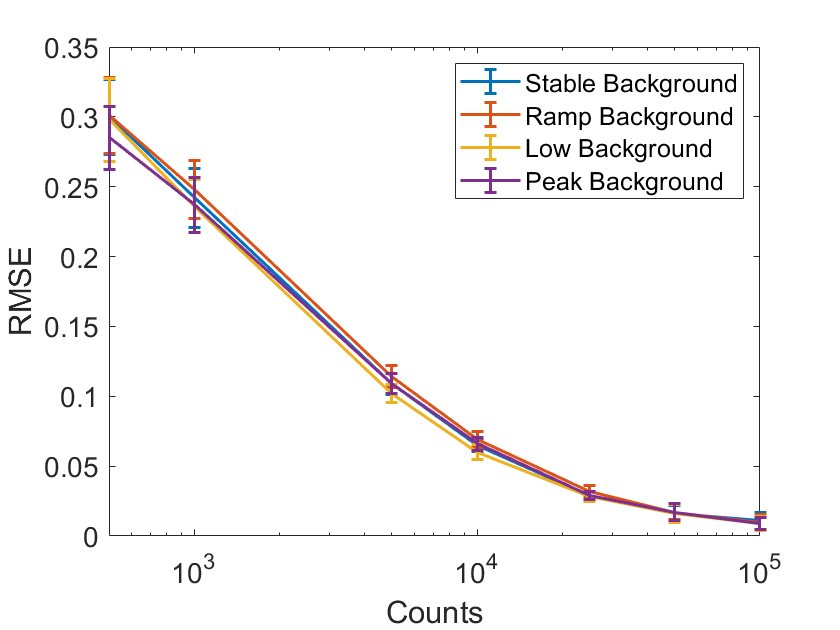}
  \caption{Mixture \#1 RMSE}
  \label{fig:rmse1}
\end{subfigure}
\begin{subfigure}[H]{0.49\textwidth}
  \includegraphics[width=\textwidth]{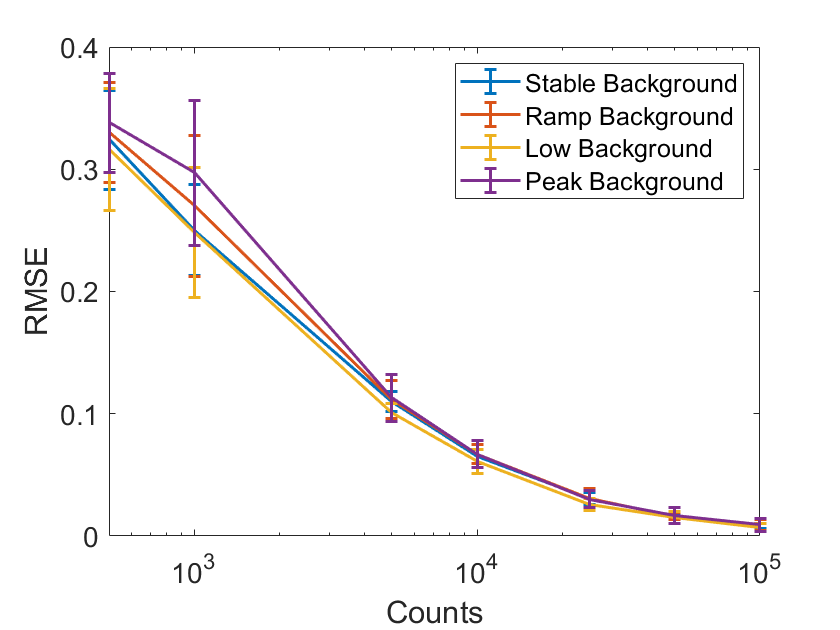}
  \caption{Mixture \#3 RMSE}
  \label{fig:rmse3}
\end{subfigure}
\begin{subfigure}[H]{0.5\textwidth}
  \includegraphics[width=\textwidth]{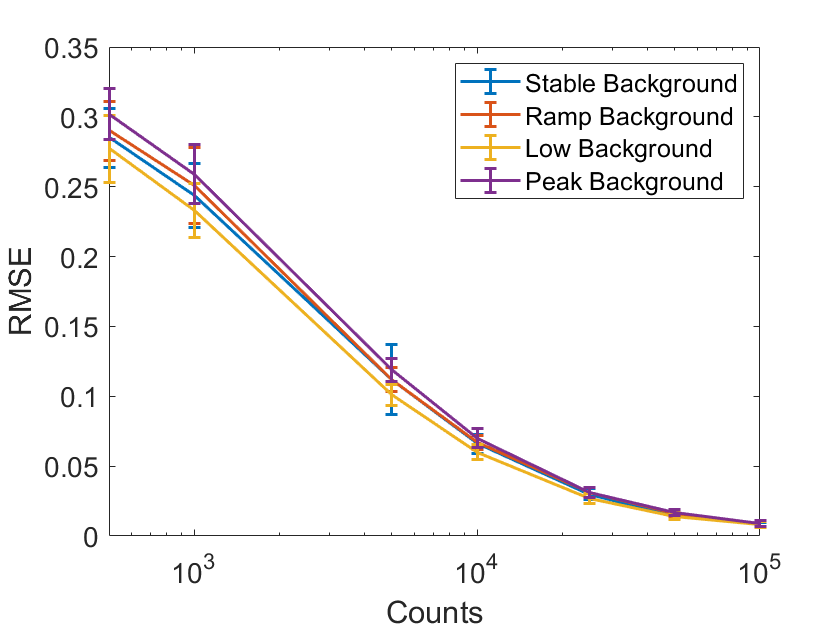}
  \caption{Mixture \#4 RMSE}
  \label{fig:rmse4}
\end{subfigure}
  \caption{RMSE comparison between different background types}
  \label{fig:rmse}
\end{figure}

Fig.~\ref{fig:rmse} shows the RMSE as a function of the increasing number of counts. The error bars shown on each line indicate the standard deviation at each count level for each background type.
These graphs show a similar trend between the RMSE for each of the background types. The unmixing algorithm correctly identifies the radionuclides when background is present, regardless of the type of background acquired at different weather conditions. 
The successfully identification is shown through each of the RMSE values being within each of the error bars, with the slight deviations potentially being caused by a weak dependence on the background spectrum. With the changing background not affecting the predicted isotope proportions significantly, a stable background spectrum in the isotope library remains an appropriate choice for isotope identification.

\section{Discussion and Conclusions}

We demonstrated that Bayesian unmixing algorithms can be a viable, computationally cost-effective approach to identify the types of radioactive materials in transit through RPMs~\cite{Altmann2020}. The overall RPM sensitivity and false alarm rate depends mostly on the detectors' intrinsic efficiency and geometric solid angle with respect to the source~\cite{Paff2016}. However, when an alarm is reported, the correct identification of SNMs is crucial to discriminate them from other sources, such as NORM or radionuclides for medical use. \added{While the minimum detectable amount of radioactivity strongly depends on system parameters, the identification performance directly depends on the unmixing algorithm and the detectors' energy resolution. \cite{Altmann2020}}. RPM performances in the field are strongly affected by changes in the radiation background, which could be due to surrounding materials or to transient atmospheric conditions. We focused on the latter problem and tested the performance of our unmixing algorithm at different background conditions. We used plastic organic scintillators to monitor the radiation background and selected background spectra during different time periods during an outdoor continuous measurement. After identifying a reference background, we have selected the background spectra corresponding to a rapidly increasing count rate, to a count rate peak, and to a count rate significantly lower than the reference. The selected background spectra mainly differed in terms of intensity, but a spectral increase in the high energy region was also found, in the presence of heavy rain, which is consistent with our expectations. The reference background should be included in the library of radionuclides, as it is expected to be present at all times. We tested the unmixing algorithm when the different background spectra were added to the mixture of radionuclides, instead of the reference background. We found that the capability of identifying the nuclides present in the mixture was not significantly affected by the presence of a background other than the reference one. We found a root-mean-square error between the true mixture fractions and the estimated ones of approximately 1.4$\%$, regardless of the type of background used, when 100,000 counts were detected. As the number of counts decreased, the RMSE increased and a slight sensitivity to the background type was observed at 500 counts. At low count regimes, not only the estimated mixture fractions can deviate from the true ones but nuclides can also be misidentified. Nonetheless, the unmixing algorithm provides an inherent uncertainty measure that can be used to determine whether a longer acquisition, with a higher signal-to-noise ratio is needed. We can conclude that the developed unmixing algorithm is robust against background changes that may occur because of varying atmospheric conditions within the 16~\degree C - 28~\degree C range. While this is not a temperature range found in harsh environments, it could be easily experienced inside an enclosure with coarse temperature control. \replaced{We are currently planning to study the identification capability of the algorithm in the presence of shielding materials, nuclides with similar energy signatures, and in dynamically changing environments where the observation noise deviates from the Poisson model.}{We are currently planning to study the sensitivity of the algorithm in the presence of shielding materials and in dynamically changing environments where the observation noise deviates from the Poisson model.} Additionally, future work can be done to account for the gain drift of the RPM detectors over time by incorporating a gain-dependent parameter in the library. This would allow for the algorithm to remain accurate without recalibrating the RPM detectors as the gain drifts over time. 

\added{
The algorithm described in this work could also be adapted to general-purpose applications dealing with sparse data to perform prompt and accurate identification of radioactive sources while being robust against changes in the background intensity.}

\section*{Acknowledgments}
This work was funded in part by the Nuclear Regulatory Commission Faculty Development Grant number 31310019M0011 and the Consortium for Verification Technology under Department of Energy National Nuclear Security Administration award number DE-NA0002534. This work was also supported by the Royal Academy of Engineering under the Research Fellowship scheme RF201617/16/31 and by the Engineering and Physical Sciences Research Council (EPSRC) Grant number EP/S000631/1 and the MOD University Defence Research Collaboration (UDRC) in Signal Processing. 

\bibliography{references}

\end{document}